\documentclass[aps,prb,reprint,showkeys,showpacs]{revtex4-1}

\usepackage{natbib}
\usepackage[utf8]{inputenc}
\usepackage{bm}
\usepackage{graphicx}
\usepackage{amsmath}
\usepackage{amsfonts}
\usepackage{xargs} 
\usepackage{color}
\usepackage{epstopdf}
\usepackage{float}
\usepackage[
colorlinks=true,
linkcolor=blue,
urlcolor=blue,
citecolor=blue,
pdftex,
bookmarks=true,
linktocpage=true,
hyperindex=true
]{hyperref}

\newcommand{\m}{\mathbf{m}}
\newcommand{\dm}{\frac{d\mathbf{m}}{dt}}
\newcommand{\HH}{\mathbf{H}}

\renewcommand{\r}[1]{{\color{black} #1}}
\renewcommand{\b}[1]{{\color{black} #1}}

\begin{document}
\title{Dynamical depinning of chiral domain walls}
\author{Simone~Moretti}
\email[Corresponding author: ]{simone.moretti@usal.es}
\affiliation{Department of Applied Physics, University of Salamanca, Plaza de los Caidos, Salamanca 37008, Spain.}
\author{Michele~Voto}
\affiliation{Department of Applied Physics, University of Salamanca, Plaza de los Caidos, Salamanca 37008, Spain.}
\author{Eduardo~Martinez}
\affiliation{Department of Applied Physics, University of Salamanca, Plaza de los Caidos, Salamanca 37008, Spain.}

\begin{abstract}
The domain wall depinning field represents the minimum magnetic field needed to move a domain wall, typically pinned by samples' disorder or patterned constrictions. Conventionally, such field is considered independent on the Gilbert damping since it is assumed to be the field at which the Zeeman energy equals the pinning energy barrier (both damping independent). Here, we analyse numerically the domain wall depinning field as function of the Gilbert damping in a system with perpendicular magnetic anisotropy and Dzyaloshinskii-Moriya interaction. Contrary to expectations, we find that the depinning field depends on the Gilbert damping and that it strongly decreases for small damping parameters.  \r{We explain this dependence with a simple one-dimensional model and we show that the reduction of the depinning field is related to the finite size of the pinning barriers and to the domain wall internal dynamics, connected to the  Dzyaloshinskii-Moriya interaction and the shape anisotropy}. 
\end{abstract}
\maketitle
\section{Introduction}
\b{Magnetic domain wall (DW) motion along ferromagnetic (FM) nanostructures has been the subject of intense research over the last decade owing to its potential for new promising technological applications~\cite{Allwood2005, Parkin2008} and for the very rich physics involved.} 
\r{A considerable effort is now focused on DW dynamics in systems with perpendicular magnetic anisotropy (PMA) which present narrower DWs and a better scalability.}  Typical PMA systems consist of ultrathin multi-layers of heavy metal/FM/metal oxide (or  heavy metal), such as ${\rm Pt/Co/Pt}$~\cite{Metaxas2007, Gorchon2014a} or ${\rm Pt/Co/AlOx}$~\cite{Moore2008a, Miron2011,Jue2016}, \b{where the FM layer has a thickness of typically $0.6-1$ nm.}  
In these systems, PMA arises mainly from interfacial interactions between the FM layer and the neighbouring layers (see Ref.~\cite{Hellman2016} and references therein). Another important interfacial effect is the Dzyaloshinskii-Moriya interaction (DMI)~\cite{Crepieux1998, Thiaville2012}, present in systems with broken inversion symmetry \r{such as Pt/Co/AlOx}. This effect gives rise to an internal in-plane field that fixes the DW chirality (the magnetization rotates always in the same direction when passing from up to down and from down to up domains) and it can  lead to a considerably faster domain wall motion~\cite{Thiaville2012} and to new magnetic patterns such as skyrmions~\cite{Rohart2013} or helices~\cite{Perez2014a}. Normally, DWs are pinned by samples' intrinsic disorder  and a minimum propagation field is needed in order to overcome such pinning energy barrier and move the DW. Such field is the DW depinning field ($H_{\rm dep}$) and it represents an important parameter from a technological point of view since a low depinning field implies less energy required to move the DW and, therefore, a energetically cheaper device.

\begin{figure}[h]
\centering
\includegraphics[width=0.35\textwidth]{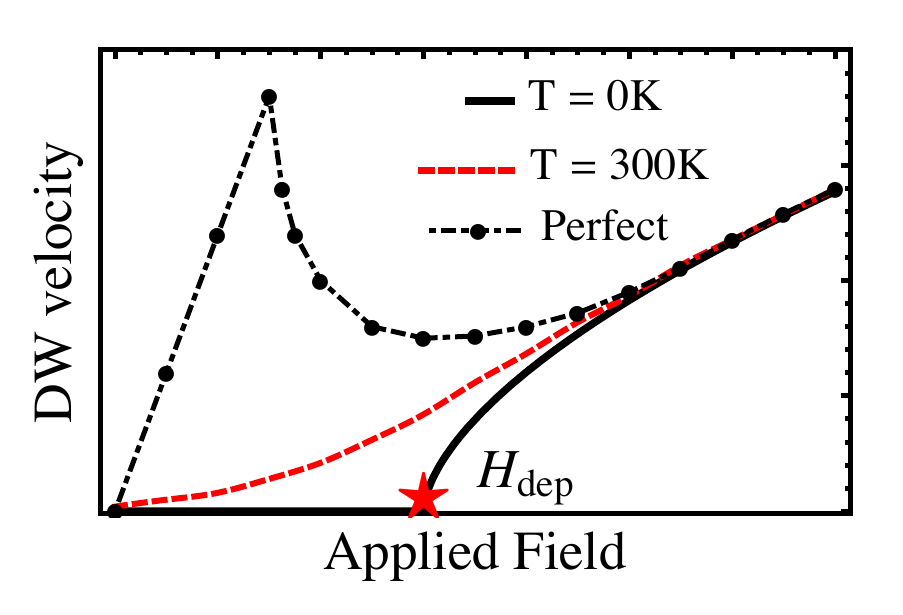}
\caption{DW velocity $vs$ applied field  as predicted by the LLG equation in a perfect system and by the {\it creep} law at $T=0$ and $T=300$K.}
\label{fig:Fig1}
\end{figure}

\r{From a theoretical point of view}, DW motion can be described by the Landau-Lifshitz-Gilbert (LLG) equation~\cite{Bertotti} which predicts, for a perfect sample without disorder, the velocity $vs$ field curve depicted in Fig.~\ref{fig:Fig1} and labelled as {\it Perfect}.  
In a disordered system, experiments have shown that a DW moves as a general one-dimensional (1D) elastic interface in a two-dimensional disordered medium~\cite{Metaxas2007, Gorchon2014a} and that it follows a theoretical velocity $vs$ driving force curve, predicted for such interfaces~\cite{Kolton2005,Agoritsas2012} (also shown in Fig.~\ref{fig:Fig1} for $T=0$ and $T=300$K). \r{Moreover},  this behaviour can be reproduced  by including disorder in the LLG equation~\cite{Martinez2007a, Voto2016a, Voto2016}. 
At zero temperature ($T=0$) the DW does not move as long as the applied field is lower than $H_{\rm dep}$, while,  at $T\neq0$, thermal activation leads to DW motion even if $H<H_{\rm dep}$ (the so called {\it creep} regime). For high fields ($H>>H_{\rm dep}$) the DW moves as predicted by the LLG equation in a perfect system. Within the creep theory, the DW is considered as a simple elastic interface and all its internal dynamics are neglected. Conventionally,  $H_{\rm dep}$ is considered independent of the Gilbert damping because it is assumed to be the field at which the Zeeman energy equals the pinning energy barrier~\cite{Metaxas2010, Franken2011} (both damping independent). Such assumption, consistently with the creep theory, neglects any effects related to the internal DW dynamics such as DW spins precession or vertical Bloch lines (VBL) formation~\cite{Yoshimura2015}. The damping parameter, for its part, represents another important parameter, which \r{controls the energy dissipation and affects the DW velocity and Walker Breakdown}~\cite{Hillebrands3}. It can be modified by doping the sample~\cite{Moore2010} or by a proper interface choice as a consequence of spin-pumping mechanism~\cite{Tserkovnyak2002}. Modifications of the DW depinning field related to changes in the damping parameter were already observed in in-plane systems~\cite{Moore2010,Pi2011} and attributed to a non-rigid DW motion~\cite{Moore2010,Pi2011}. \r{Oscillations of the DW depinning field due to the internal DW dynamics were also experimentally observed in in-plane similar systems~\cite{Thomas2006}.} \r{Additional dynamical effects in soft samples, such as DW boosts in current induced motion, were numerically predicted and explained in terms of DW internal dynamics and DW transformations~\cite{Yuan2015,Yuan2015a}.  } 

Here, we numerically analyse the DW depinning field in a system with PMA and DMI as function of the Gilbert damping. \r{We observe a reduction of $H_{\rm dep}$ for low damping and we explain this behaviour by adopting a simple 1D model. We show that the effect is due to the finite size of pinning barriers and to the DW internal dynamics, related to the DMI and shape anisotropy fields.} This article is structured as follows: in Section~\ref{sec:methods} we present the simulations method, the disorder implementation and the $H_{\rm dep}$ calculations. The main results are outlined and discussed in Section~\ref{sec:results}, where we also present the 1D model.  Finally, the main conclusions of our work are summarized in Section~\ref{sec:conclusions}.

\section{Micromagnetic simulations}
\label{sec:methods}
\begin{figure}[h]
\includegraphics[width=0.45\textwidth]{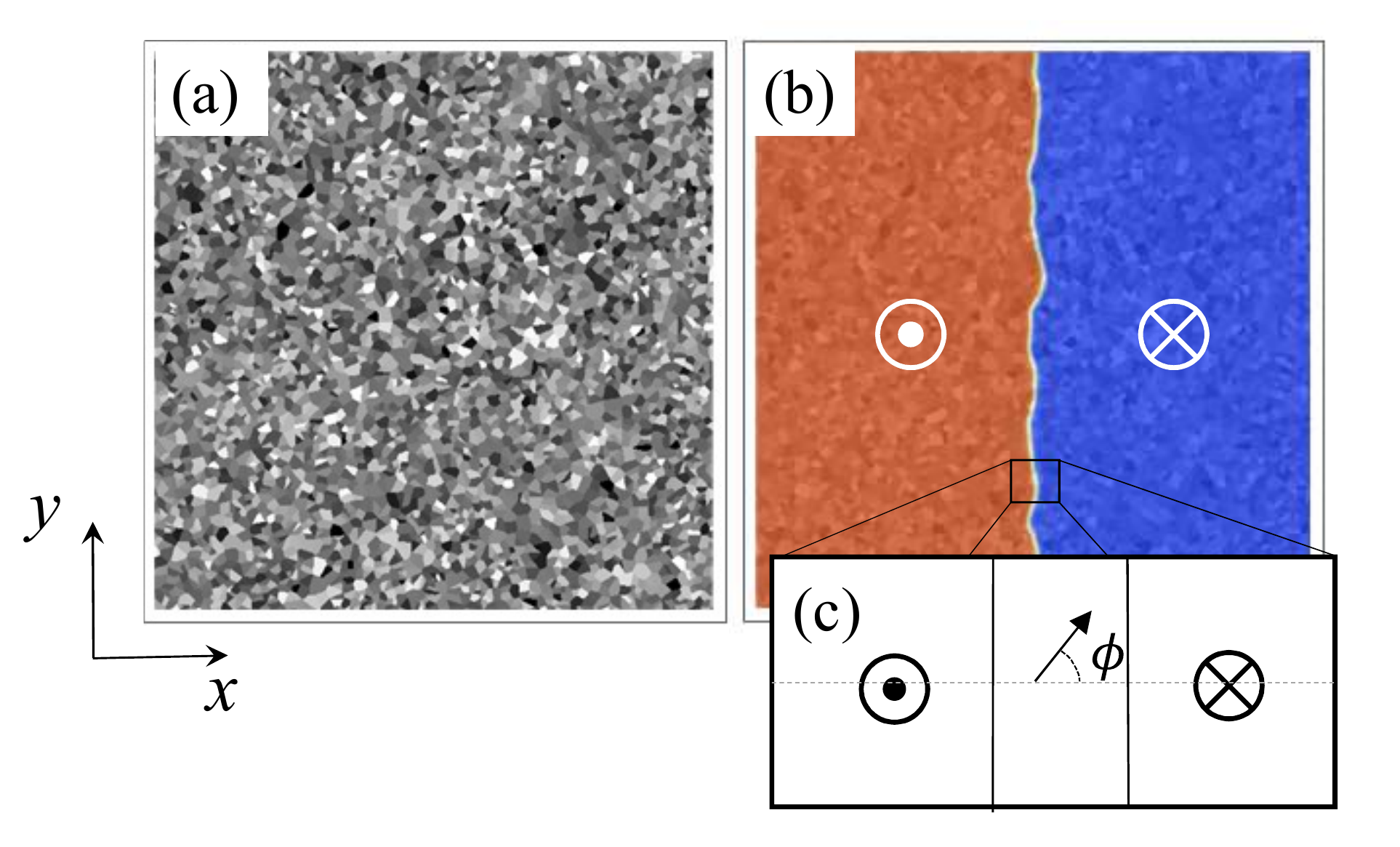}
\caption{(a) Grains structure obtained by Voronoi tassellation. (b) Initial DW state. (c) Sketch of the internal DW angle $\phi$. }
\label{fig:Fig2}
\end{figure}

We consider a sample of dimensions $(1024~\times~1024~\times~0.6)~{\rm nm^3}$ with periodic boundary conditions along the $y$ direction, in order to simulate an extended thin film. Magnetization dynamics is analysed by means of the LLG equation~\cite{Bertotti}:
\begin{eqnarray}
\dm = -\frac{\gamma_0}{1+\alpha^2}\left(\m\times\HH_{\rm eff}\right)-\frac{\gamma_0\alpha}{1+\alpha^2}\left[\m\times\left(\m\times\HH_{\rm eff}\right)\right]\, ,\nonumber\\
\label{eq:LLG}
\end{eqnarray}
\r{where $\m(\mathbf{r},t)=\mathbf{M}(\mathbf{r},t)/M_s$ is the normalized magnetization vector, with $M_s$ being the saturation magnetization. $\gamma_0$ is the gyromagnetic ratio and $\alpha$ is the Gilbert damping. $\HH_{\rm eff}=\HH_{\rm exch}+\HH_{\rm DMI}+\HH_{\rm an}+\HH_{\rm dmg}+H_z\hat{\mathbf{u}}_z$ is the effective field, including the exchange, DMI, uniaxial anisotropy, demagnetizing and external field contributions~\cite{Bertotti} respectively.} Typical PMA samples parameters are considered: $A=17\times 10^{-12}\ {\rm J/m}$, $M_s=1.03\times 10^6 \ {\rm A/m}$, $K_u=1.3\times 10^6\ {\rm J/m^3}$ and $D=0.9\ {\rm mJ/m^2}$, where $A$  is the exchange constant, $D$ is the DMI constant and $K_u$ is the uniaxial anisotropy constant. Disorder is taken into account by dividing the sample into grains by Voronoi tessellation~\cite{Vansteenkiste2014,Leliaert2014}, as shown in Fig.~\ref{fig:Fig2}(a). In each grain the micromagnetic parameters $\{M_{s}, D_{c}, K_{u}\}$ change in a correlated way in order to mimic a normally distributed thickness~\cite{Hrabec2016}:
\begin{eqnarray}
t_G=N(t_0,\delta)\rightarrow\left\{
\begin{array}{ccc}
M_G&=&(M_{s}t_G)/t_0\\
K_G&=&(K_{u}t_0)/t_G\\
D_G&=&(D_{c}t_0)/t_G\\
\end{array}\right.\, ,
\label{eq:dis}
\end{eqnarray}
where the subscript $G$ stands for grain, $t_0$ is the average thickness ($t_0=0.6{\rm nm}$) and $\delta$ is the standard deviation of the thickness normal distribution. 
The sample is discretized in cells of dimensions $(2\times 2\times 0.6){\rm nm^3}$, smaller than the exchange length $l_{ex}\sim 5{\rm nm}$.  Grain size is GS=$15\ {\rm nm}$, reasonable for these materials, while the thickness fluctuation is  $\delta=7\%$. 
Eq.~(\ref{eq:LLG}) is solved by the finite difference solver MuMax~3.9.3~\cite{Vansteenkiste2014}.

A DW is placed and relaxed at the center of the sample as depicted in Fig.~\ref{fig:Fig2}(b). $H_{\rm dep}$ is calculated by applying \r{a sequence} of fields and running the simulation, \b{for each field, until the DW is expelled from the sample, or until the system has reached an equilibrium state (i.e. the DW remains pinned): $\tau_{\rm max}<\epsilon (\alpha)$}. $\tau_{\rm max}$ indicates the maximum torque, which rapidly decreases when the system is at equilibrium.  It only depends on the system parameters and damping. For each value of $\alpha$, we choose a specific threshold, $\epsilon(\alpha)$, in order to be sure that we reached an equilibrium state (see Supplementary Material~\footnote{See Appendices} for more details). The simulations are repeated for $20$ different disorder realizations. Within this approach, $H_{\rm dep}$ corresponds to the minimum field needed to let the DW propagate freely through the whole sample. In order to avoid boundaries effects, the threshold for complete depinning is set to $\langle m_z \rangle >0.8$, 
 \r{where $\langle m_z\rangle$ is averaged over all the realizations, i.e. $\langle m_z\rangle=\sum_{i=1}^N\langle m_z\rangle_i/N$, where $N=20$ is the number of realizations. We checked that, in our case, this definition of $H_{\rm dep}$ coincides with taking $H_{\rm dep}={\rm Max}\{H_{\rm dep}^i\}$, with $H_{\rm dep}^i$ being the depinning field of the single realization.}
\r{In other words, $H_{\rm dep}$ corresponds to the minimum field needed to depin the DW from any possible pinning site considered in the 20 realizations~\footnote{This definition is preferred over the average of $H_{\rm dep}^i$ since it is more independent on the sample size. In fact, by increasing the sample dimension along the $x$ direction, we increase the probability of finding the highest possible $h_j$ in the single realization and the average of $H_{\rm dep}^i$ will tend to the maximum. }. }

\r{Following this strategy,} the DW depinning field is numerically computed with two different approaches:\\ $(1)$ by {\it Static} simulations, which neglect any precessional dynamics by solving
\begin{eqnarray}
\dm = -\r{\frac{\gamma_0\alpha}{1+\alpha^2}}\left[\m\times\left(\m\times\HH_{\rm eff}\right)\right]\, .
\label{eq:LLGrelax}
\end{eqnarray}
This is commonly done when one looks for a minimum of the system energy and it corresponds to the picture in which $H_{\rm dep}$ simply depends on the balance between Zeeman and pinning energies.\footnote{This is solved by the {\it Relax} solver of MuMax with the assumption $\alpha/(1+\alpha^2)=1$.}\\
$(2)$ by {\it Dynamic} simulations, which include precessional dynamics by solving the full  Eq.~(\ref{eq:LLG}). This latter method corresponds to the most realistic case. 
\r{Another way to estimate the depinning field is to calculate the DW velocity $vs$ field curve at $T=0$ and look for minimum field at which the DW velocity is different from zero. For these simulations we use a moving computational region and we run the simulations for $t=80{\rm ns}$ (checking that longer simulations do not change the DW velocity, meaning that we reached a stationary state). This second setup requires more time and the calculations are repeated for only $3$ disorder realizations.}

Using these methods, the depinning field $H_{\rm dep}$ is calculated for different damping parameters $\alpha$.  

\section{Results and discussion}
\label{sec:results}
\subsection{Granular system}
\begin{figure}[h]
\centering
\includegraphics[width=0.36\textwidth]{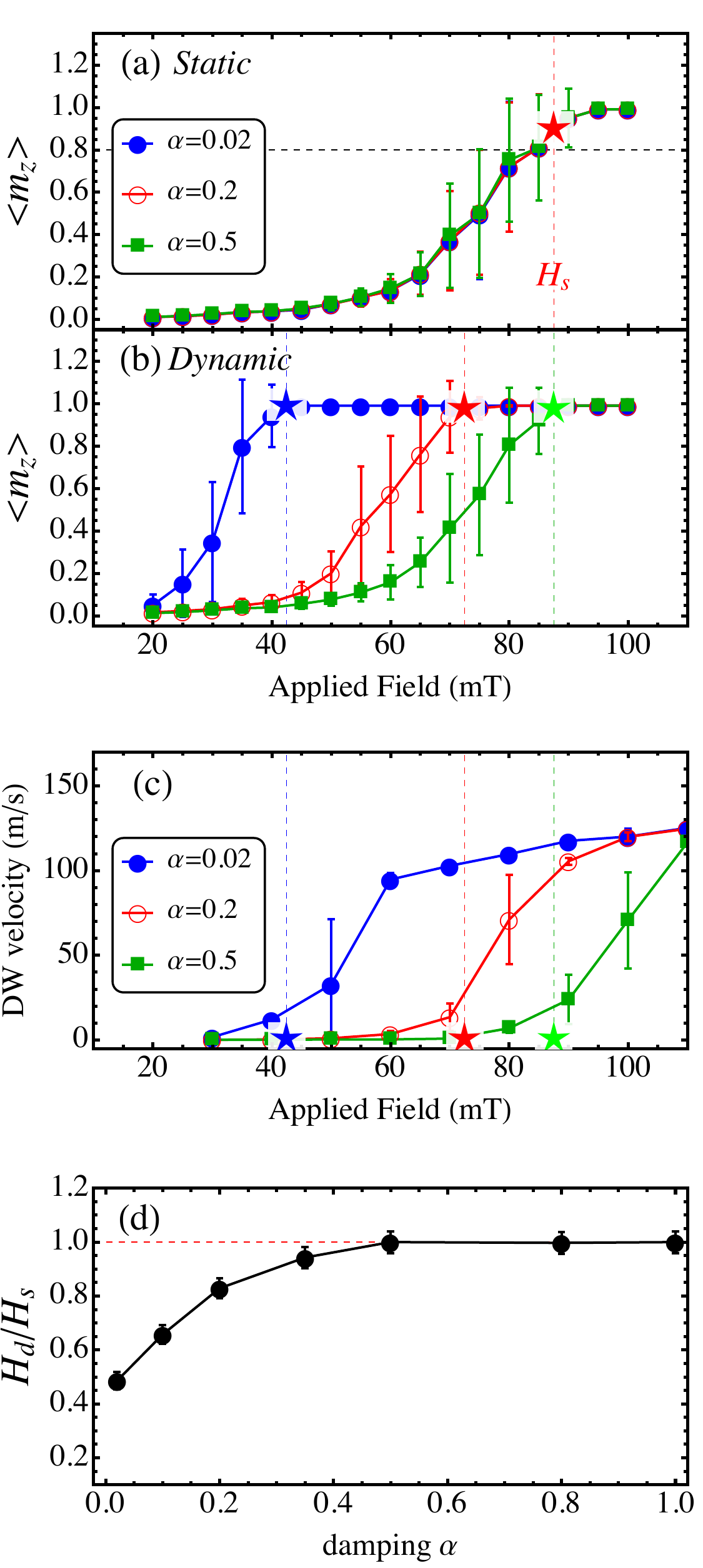}
\caption{Average $\langle m_z\rangle$ as function of applied field for different damping parameters for the (a){\it Static} simulations and (b){\it Dynamic} simulations. (c) DW velocity vs applied field for different damping. (d) Dynamical depinning field, normalized to $H_s$, as function of damping.}
\label{fig:Fig3}
\end{figure}

Our first result is shown in Fig.~\ref{fig:Fig3}(a)-(b), which depicts the final average magnetization $\langle m_z\rangle$ as function of the applied field for different damping parameters. 
\r{In the {\it Static} simulations (Fig.~\ref{fig:Fig3}(a)) $H_{\rm dep}$ does not depend on damping, so that a static depinning field can be defined.  Conversely, in the {\it Dynamic} simulations (Fig.~\ref{fig:Fig3}(b)), \b{$H_{\rm dep}$ decreases for low damping parameters}. The depinning field is indicated by a star in each plot and the static depinning field is labelled as $H_s$.} The same result is obtained by calculating $H_{\rm dep}$ from the DW velocity $vs$ applied field plot, shown in Fig.~\ref{fig:Fig3}(c). The stars in  Fig.~\ref{fig:Fig3}(c) correspond to the depinning fields calculated in the previous simulations and they are in good agreement with the values predicted by the velocity $vs$ field curve. \r{The dynamical depinning field $\mu_0 H_d$, normalized to the static depinning field  $\mu_0 H_s=(87\pm 1 ){\rm mT}$, with $\mu_0$ being the vacuum permeability, is shown in Fig.~\ref{fig:Fig3}(d) as function of the damping parameter $\alpha$.} $H_d$ saturates for high damping (in this case $\alpha\geq0.5$) while it decreases for low damping until $H_d/H_s\sim0.4$ at $\alpha=0.02$. This reduction must be related to the precessional term, neglected in the static simulations. The same behaviour is observed with different grain sizes (GS=$5$ and $30$ nm) and with a different disorder model, consisting of a simple variation of the $K_u$ module in different grains. This means that the effect is not related to the grains size or to the particular disorder model we used. 

\begin{figure}[h]
\centering
\includegraphics[width=0.38\textwidth]{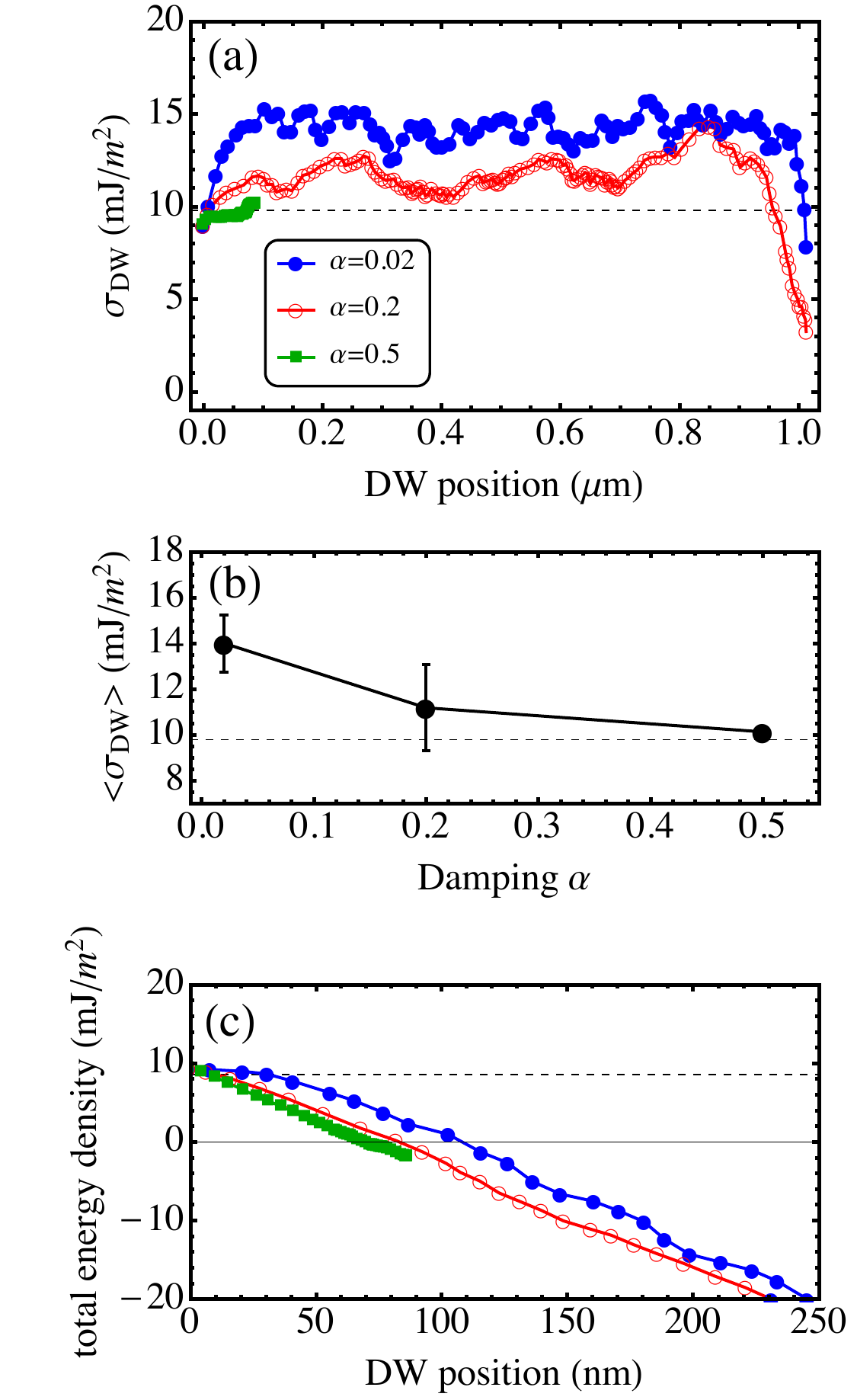}
\caption{(a) DW energy density as function of DW position for different damping. The final drop corresponds to the expulsion of the DW. (b) Average DW density as funciton of damping. Dashed line represents the analytical value $\sigma_{\infty}\sim 10\ {\rm mJ/m^2}$. \r{(c) Total energy density of the system as function of DW position for different damping parameters. } }
\label{fig:Fig5}
\end{figure}

Additionally, Fig.~\ref{fig:Fig5} represents the DW energy~\footnote{The DW energy is calculated as the energy of the system with the DW minus the energy of the system without the DW (uniform state). The profile is obtained by moving the DW with an external applied field and then subtracting the Zeeman energy. } as function of DW position and damping parameter \r{for $\mu_0 H_z=70$ mT}. At high damping, the average DW energy density converges to $\sigma_{\infty}\sim 10\ {\rm mJ/m^2}$, in good agreement with the analytical value $\sigma_0=4\sqrt{AK_0}-\pi D=10.4\ {\rm mJ/m^2}$, where $K_0$ is the effective anisotropy $K_0=K_u-\mu_0 M_s^2/2$. On the contrary, for low damping, the DW energy increases up to $\sigma (0.02)\sim 14\ {\rm mJ/m^2}$.  This increase,  related to DW precessional dynamics, reduces the effective energy barrier and helps the DW to overcome the pinning barriers. \r{Fig.~\ref{fig:Fig5}(c) shows the total energy of the system (including Zeeman). As expected~\cite{Wang2009}, the energy decreases as the DW moves.} 

\begin{figure*}[t]
\includegraphics[width=0.9\textwidth]{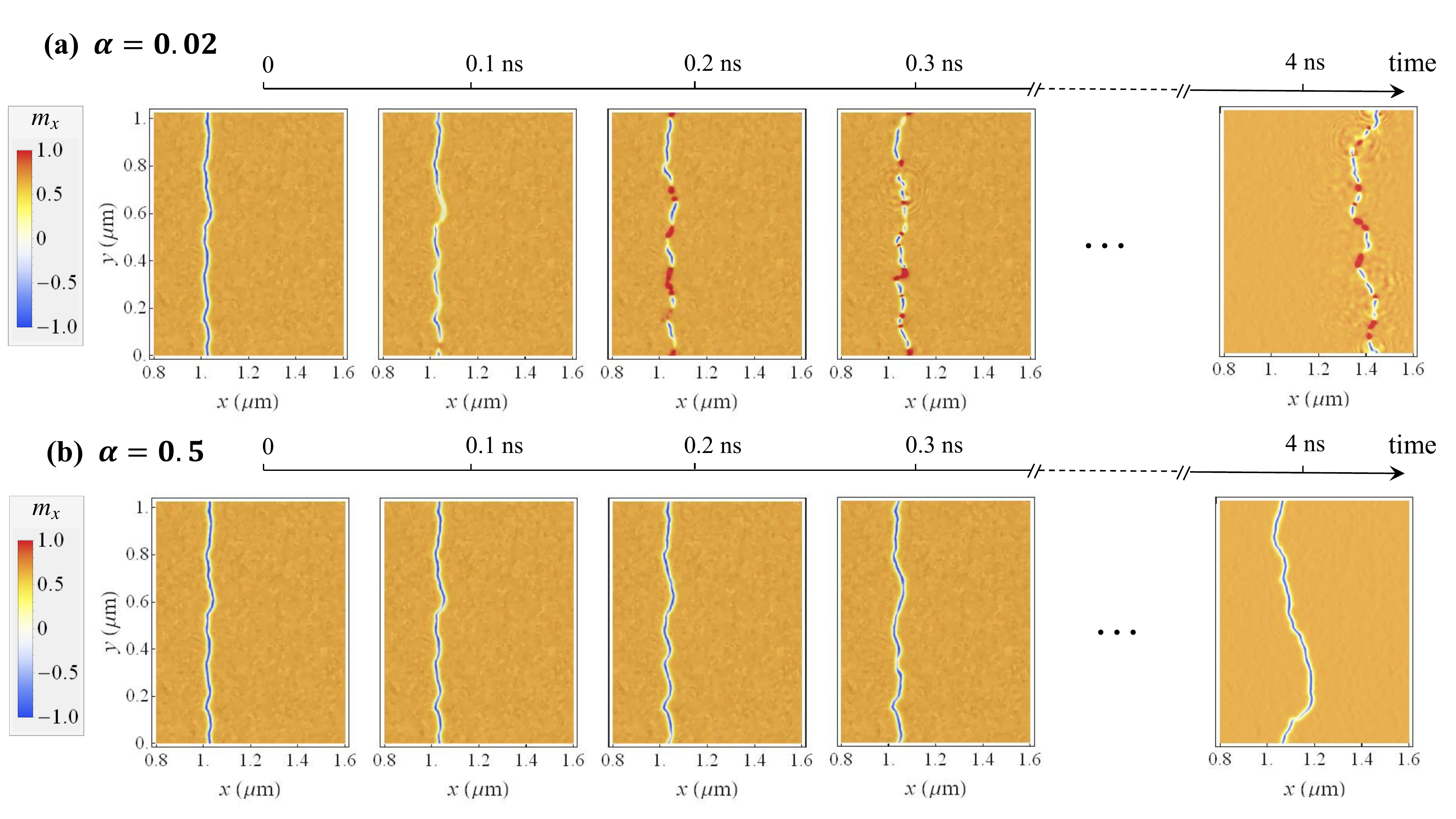}
\caption{(a) Snapshots of the magnetization dynamics at subsequent instants under $\mu_0 H_z=70 {\rm mT}$, for two different damping: (a) $\alpha=0.02$ and (b) $\alpha=0.5$. The grains pattern, and therefore the energy barrier, is the same for both cases. \b{In order to let the DW move across more pinning sites, these simulations were performed on a larger sample with $L_x=2048$ nm. }}
\label{fig:Fig4}
\end{figure*}

\r{Finally}, Fig.~\ref{fig:Fig4} shows the DW motion as function of time for $\alpha=0.02$ and $\alpha=0.5$, along the same grain pattern (and therefore along the same pinning barriers). The applied field is $\mu_0H_z=70 {\rm mT}$, which satisfies $H_d(0.02)<H_z<H_d(0.5)$. The initial DW configuration is the same but, for $\alpha=0.02$, VBL start to nucleate and the DW motion is much more turbulent (see Supplementary Material~\cite{Note1} for a movie of this process). At $t=4\ {\rm ns}$ the DW has reached an equilibrium position for $\alpha=0.5$, while it has passed through the (same) pinning barriers for $\alpha=0.02$. \b{Thus, one might think that the reduction of the depinning field could be related to the presence of VBL and their complex dynamics~\cite{Yoshimura2015}.}
Further insights about this mechanism are given by analysing the DW depinning at a single energy barrier as described in the next subsection.\\

\subsection{Single barrier}
\begin{figure}[h]
\centering
\includegraphics[width=0.4\textwidth]{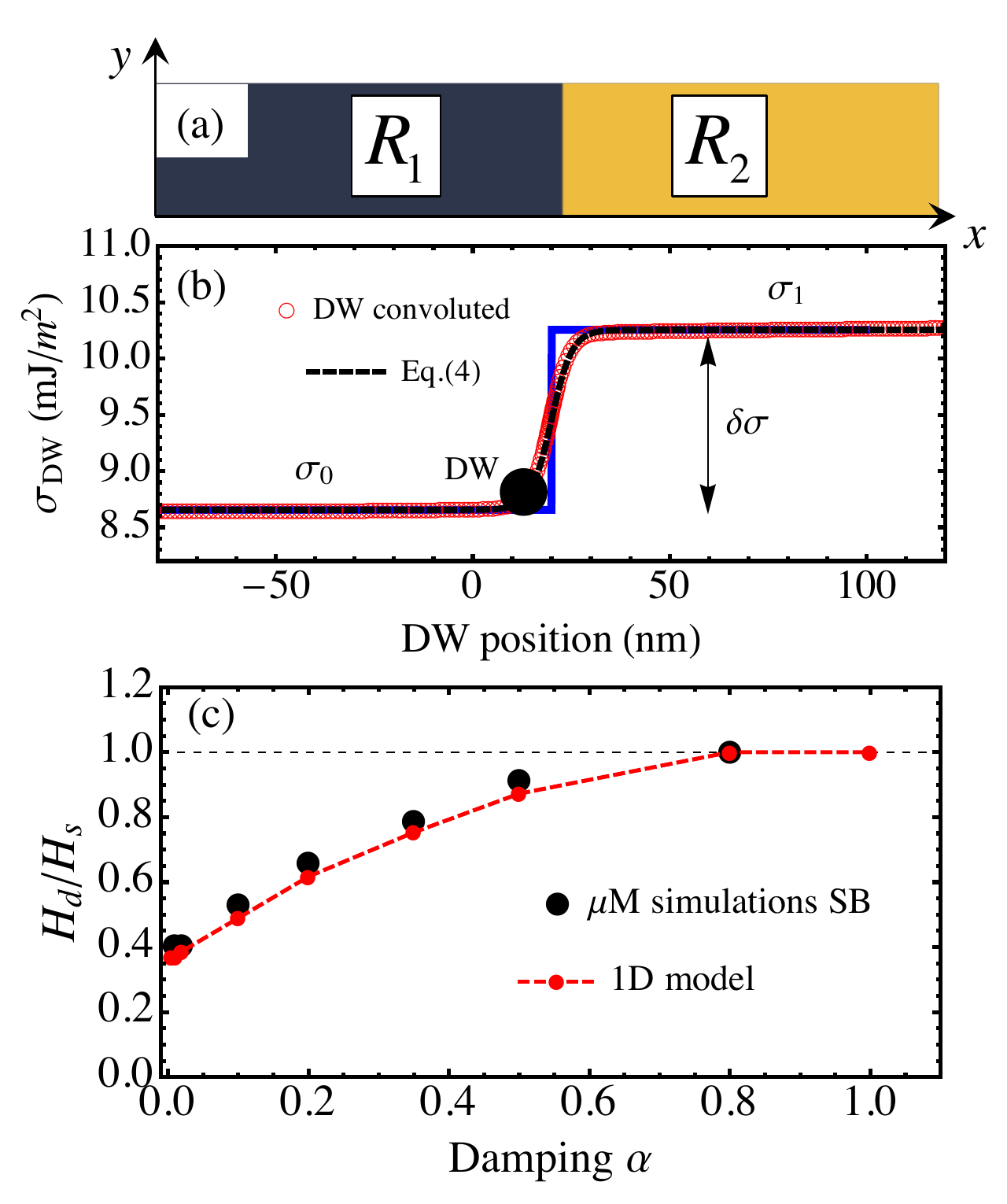}
\caption{(a) Sketch of the two regions implemented for the single barrier (SB) micromagnetic simulations. (b) DW energy as function of DW position along the strip. Blue solid line represents the analytical value, red points the DW convoluted energy (due to the finite size of the DW) while black dashed line a fit using Eq.~\ref{eq:DW_en}. (c) Dynamical depinning field, normalized to the static depinning field, for the single barrier simulations as function of damping, obtained from full micromagnetic simulations and the 1D model.}
\label{fig:Fig6}
\end{figure}

In order to understand how the DW precessional dynamics reduces $H_{\rm dep}$, we micromagnetically analysed the DW depinning from a single barrier as sketched in Fig.~\ref{fig:Fig6}. We considered a strip of dimensions $(1024\times256\times0.6){\rm nm^3}$ and we divided the strip into two regions, $R_1$ and $R_2$, which are assumed to have a thickness of $t_{1}=0.58$ and $t_{2}=0.62$ nm respectively. Their parameters vary accordingly (see Sec.~\ref{sec:methods}), generating the DW energy barrier ($\delta\sigma$) shown in Fig.~\ref{fig:Fig6}(b). \r{A DW is placed and relaxed just before the barrier.} The finite size of the DW ($\pi\Delta_{\rm DW}\sim 15\ {\rm nm}$, with $\Delta_{\rm DW}$ being the DW width parameter) smooths the abrupt energy step and, in fact, the energy profile can be successfully fitted by using the Bloch profile~\cite{Hillebrands3}
\begin{eqnarray}
\sigma_{\rm DW}&=&\sigma_0+\nonumber\\
&+&\left(\frac{\delta\sigma}{2}\right)\left\{1+\cos\left(2\arctan\left[\exp\left(\frac{x_0-x}{\Delta_{\rm DW}}\right)\right]\right)\right\}\, ,\nonumber\\
\label{eq:DW_en}
\end{eqnarray}
where $x_0=20\ {\rm nm}$ is the step position, while $\sigma_0$ and $\sigma_1$ are the DW energies at the left and right side of the barrier as represented in Fig.~\ref{fig:Fig6}(b). This means that the pinning energy barrier has a spatial extension which is comparable to the DW width. 
\b{By performing the same static and dynamic simulations, we obtain a static depinning field of $\mu_0H_s=120\ {\rm mT}$ and, when decreasing the damping parameter, we observe the same reduction of the depinning field as in the granular system (see Fig.~\ref{fig:Fig6}(c)). In this case the DW behaves like a rigid object whose spins precess coherently and no VBL nucleation is observed.  Hence,  $H_{\rm dep}$ reduction does not depend directly on  the presence of VBL but on the more general mechanism of spins' precession already present in this simplified case.}\\
\r{Nevertheless, an important characteristic of these single barrier simulations is that the barrier is  localized and it has a finite size which is of the order of the DW width. Note that the same holds for the granular system: despite a more complex barrier structure, the dimension of the single barrier between two grains has the size of the DW width.} \\

\begin{figure*}[t]
\centering
\includegraphics[width=0.9\textwidth]{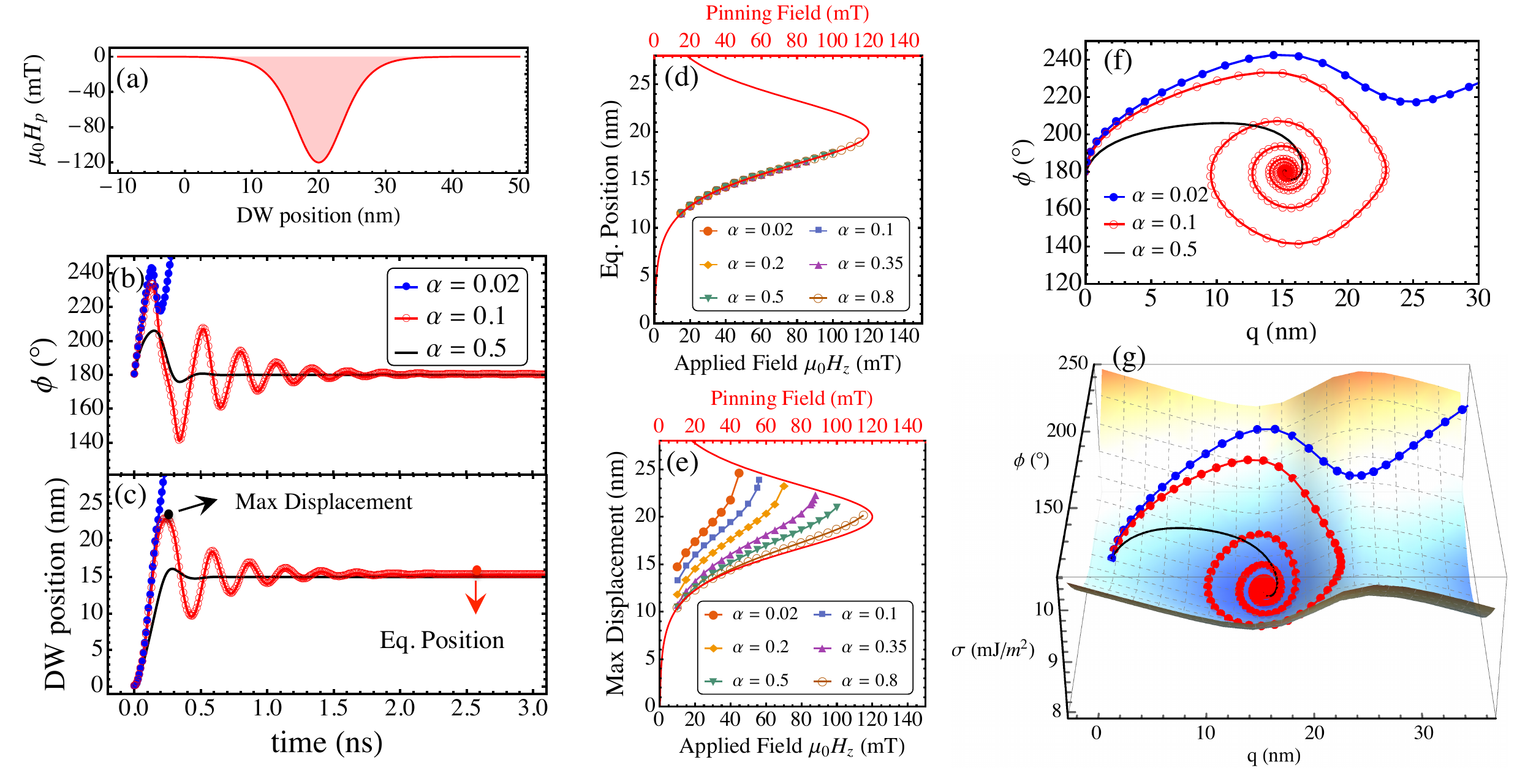}
\caption{\r{(a) Pinning field obtained from Eq.~(\ref{eq:Hp}) as function of DW position.  DW position  internal angle $\phi$ as function of time for different damping parameter and $\mu_0H_z=55$ mT. (c) DW position $q$ as function of time for different damping and $\mu_0H_z=55$ mT. (d) Equilibrium position as function of applied field for different damping. (e) Maximum DW displacement as function of the applied field for different damping. (f) DW coordinates $\{q,\phi\}$ for $\mu_0H_z=55$ mT and different damping. (g) DW coordinates $\{q,\phi\}$ inside the energy landscape: $\sigma=\sigma_{\rm DW}(q,\phi)-2\mu_0M_sH_zq$.} }
\label{fig:Fig7}
\end{figure*}

Thus, in order to understand the interplay between the DW precessional dynamics and the finite size of the barrier,  we considered a 1D collective-coordinate model with a localized barrier. The 1D model equations, describing the dynamics of the DW position $q$ and the internal angle $\phi$ (sketched in Fig.~\ref{fig:Fig2}(c)), are given by~\cite{Martinez2007a}

\begin{eqnarray}
(1+\alpha^2)\dot\phi&=&\gamma_0 [(H_z+H_p(q))\nonumber\\
&&-\alpha \underbrace{\left(H_K\frac{\sin 2\phi}{2}-\frac{\pi}{2}H_{\rm DMI}\sin\phi\right)}_{H_{\rm int}(\phi)}]\, ,\label{phi}\\
(1+\alpha^2)\frac{\dot q}{\Delta_{\rm DW}}&=&\gamma_0\left[ \alpha(H_z+H_p(q))\right.\nonumber\\
&&+\left.\left(H_K\frac{\sin 2\phi}{2}-\frac{\pi}{2}H_{\rm DMI}\sin\phi\right)\right]\, ,\label{q}
\end{eqnarray}
where $H_K=M_s N_x$ is the shape anisotropy field, favouring Bloch walls, with $N_x=t_0\log 2/(\pi\Delta_{DW})$~\cite{Tarasenko1998} being the DW demagnetizing factor along the $x$ axis. $H_{\rm DMI}=D/(\mu_0M_s\Delta_{DW})$ is the DMI field.  $H_{\rm int}(\phi)$ represents the internal DW field, which includes DMI and shape anisotropy. $H_{\rm int}$ favours Bloch ($\phi=\pm\pi/2$) or Néel wall ($\phi=0$ or $\phi=\pi$) depending on the relative strength of $H_K$ and $H_{\rm DMI}$. In our system, the DMI dominates over shape anisotropy since $\mu_0H_{\rm DMI}\sim170$ mT while $\mu_0H_K\sim30$ mT. Hence, the DW equilibrium angle is $\phi=\pi$ ($\phi=0$ or $\phi=\pi$ additionally depends on the sign of the DMI).  $H_p(q)$ is the DW pinning field, obtained from the DW energy profile (Eq.~(\ref{eq:DW_en})) as follows: the maximum pinning field is taken from the static simulations while the shape of the barrier is taken as the normalized DW energy gradient (see Supplementary Material~\cite{Note1} for more details),
\begin{eqnarray}
&&H_p(q)=H_s\left(\frac{\partial\sigma_{\rm DW}(x)}{\partial x}\right)_N=\nonumber\\
&&=2H_s\frac{\exp\left(\frac{x_0-q}{\Delta_{DW}}\right)\sin\left[2\arctan\left(\exp\left(\frac{x_0-q}{\Delta_{DW}}\right)\right)\right]}{1+\exp\left(\frac{2(x_0-q)}{\Delta_{DW}}\right)}\, .
\label{eq:Hp}
\end{eqnarray}
The corresponding pinning field is plotted in Fig.~\ref{fig:Fig7}(a).~\footnote{The same results are obtained with a Gaussian barrier, meaning that the key point is the finite size of the barrier rather than its shape.}\\
The results for the dynamical $H_{\rm dep}$, obtained with this modified 1D model, are plotted in Fig.~\ref{fig:Fig6}(c) and they show a remarkable agreement with the single barrier micromagnetic simulations. This indicates that the main factors responsible for the reduction of $H_{\rm dep}$ are already included in this simple 1D model. \r{Therefore, additional insights might come from analysing the DW dynamics within this 1D model. Fig.~\ref{fig:Fig7}(b) and (c) represents the DW internal angle $\phi$ and the DW position $q$ as function of time for different damping. The plots are calculated with $\mu_0H_z=55$ mT which satisfies $H_{\rm dep} (0.02)<H_z<H_{\rm dep}(0.1)<H_{\rm dep}(0.5)$. As shown in Fig.~\ref{fig:Fig7}(b) and (c),  below the depinning field ($\alpha=0.1$, $\alpha=0.5$), both the internal angle and the DW position oscillate before reaching the same final equilibrium state. However, the amplitude of these oscillations (the maximum displacement) depends on the damping parameter. 
\r{ Fig.~\ref{fig:Fig7}(d) shows the final equilibrium position as function of the applied field for different damping.} \r{The equilibrium position is the same for all damping and it coincides with the position at which $H_z=H_p(q)$. Conversely, the maximum displacement, shown in Fig.~\ref{fig:Fig7}(e), strongly  increases for low damping parameters. For applied field slightly smaller than the depinning field,  the DW reaches the boundary of the pinning barrier, meaning that a further increase of the field is enough to have a maximum displacement higher than the barrier size and depin the DW. In other words, the decrease of the depinning field, observed in the single barrier simulations, is due to DW oscillations that depend on $\alpha$ and that can be larger than the barrier size, leading to DW depinning for lower field. 
The DW dynamics and the depinning mechanism are further clarified in Fig.~\ref{fig:Fig7}(f) and Fig.~\ref{fig:Fig7}(g).   Fig.~\ref{fig:Fig7}(f) represents the DW coordinates $\{q,\phi \}$ for $\mu_0 H_z=55$ mT and different damping. Before reaching the common equilibrium state, the DW  moves in orbits (in the $\{q,\phi\}$ space) whose radius depends on the damping parameter. For $\alpha=0.5$ (black line) the DW rapidly collapse into the final equilibrium state. Conversely, for $\alpha=0.1$ (red open circles), the DW orbits around the equilibrium state before reaching it. If the radius of the orbit is larger than the barrier size the DW gets depinned, as in the case of $\alpha=0.02$ (blue full circles). This mechanism is also represented in Fig.~\ref{fig:Fig7}(g), where the DW orbits are placed in the energy landscape. The energy is calculated as $\sigma (q,\phi)=\sigma_{\rm DW }(q,\phi)-2\mu_0M_s H_z q$, where $\sigma_{\rm DW}$ is given by Eq.~(\ref{eq:DW_en}).  Fig.~\ref{fig:Fig7}(g) shows that the equilibrium state corresponds to the new minimum of the energy landscape. Furthermore, it confirms that the applied field is below the static depinning field, at which the pinning barrier would have been completely lifted.   Nevertheless, while reaching the equilibrium state, the DW moves inside the energy potential and, if the radius of the orbit is larger than the barrier size, the DW can overcome the pinning barrier, as shown for $\alpha=0.02$ in Fig.~\ref{fig:Fig7}(g).}}   
 
\begin{figure}[h]
\centering
\includegraphics[width=0.36\textwidth]{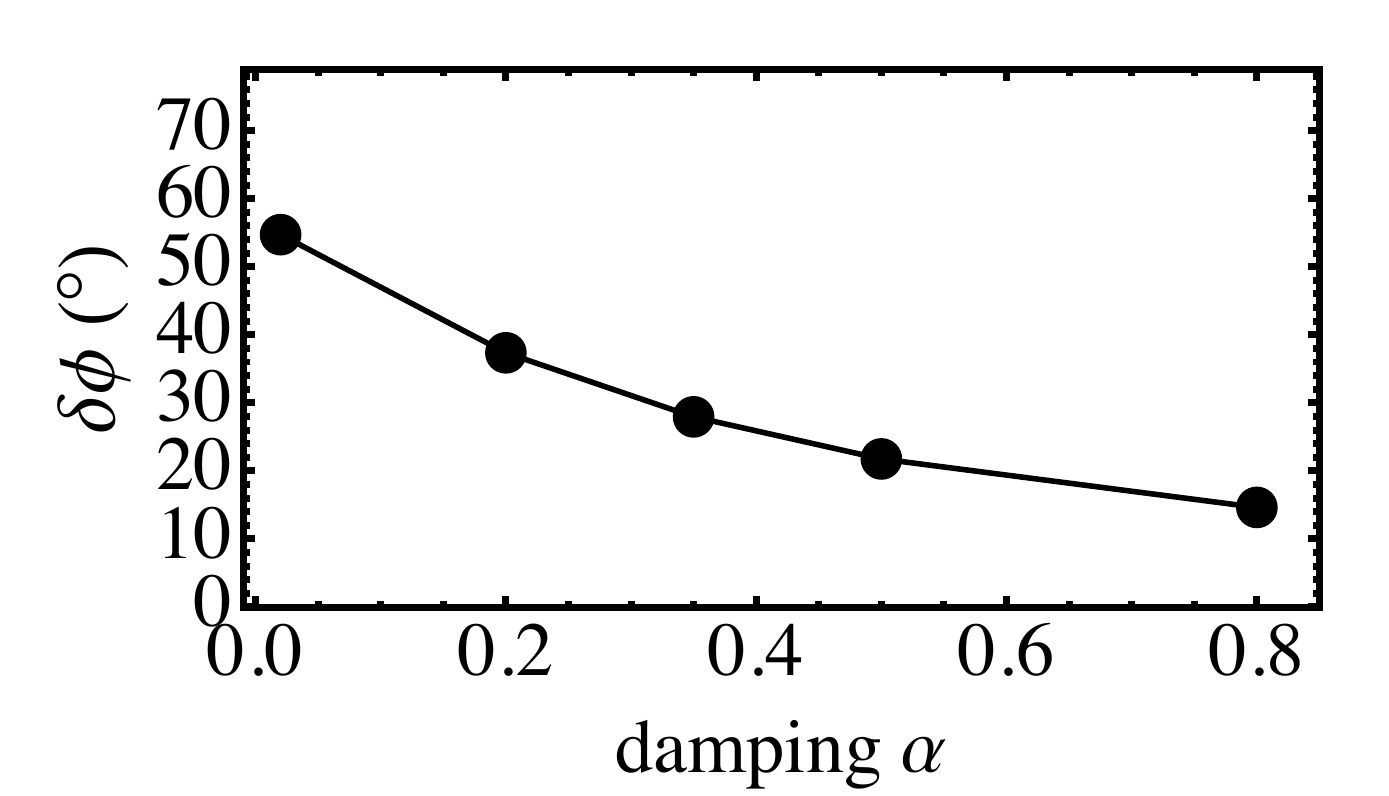}
\caption{ Maximum deviation of $\phi$ from its equilibrium position as function of damping. }
\label{fig:Fig8}
\end{figure}

\b{At this point we need to understand why the amplitude of the DW oscillations depends on damping. 
By solving Eq.~(\ref{phi}) and Eq.(\ref{q}) for the equilibrium state ($\dot{q}=0$, $\dot{\phi}=0$) we obtain 
\begin{eqnarray}
\dot{q}=0\Rightarrow |H_p(q)|&=&H_z+\frac{H_{\rm int}(\phi)}{\alpha}\nonumber\\ 
&\approx & H_z-\frac{\pi}{2}\frac{H_{\rm DMI}}{\alpha}\sin\phi\, ,
\label{eq:q0}\\
\dot{\phi}=0\Rightarrow |H_p(q)|&=&H_z-\alpha H_{\rm int}(\phi)\nonumber\\
&\approx & H_z+\alpha\frac{\pi}{2} H_{\rm DMI}\sin\phi\, ,
\label{eq:phi0}
\end{eqnarray}
since $\mu_0 H_{\rm DMI} \gg \mu_0 H_{K}$ and, therefore, $H_{\rm int}\approx -(\pi/2) H_{\rm DMI}\sin\phi$. These equations have a single common solution which corresponds to $|H_p(q)|=H_z$ and $\phi=\phi_0=\pi$ (at which $H_{\rm int}(\pi)=0$). \r{However, at \r{$t=0$}, the DW starts precessing under the effect of the applied field and, if $\phi\neq\pi$ when $|H_p(q)|=H_z$, the DW does not stop at the final equilibrium position but it continues its motion, as imposed by Eq.~(\ref{eq:q0}) and (\ref{eq:phi0}). 
In other words, the DW oscillations in Fig.~\ref{fig:Fig7}(b) are given by oscillations of the DW internal angle $\phi$, around its equilibrium value $\phi_0=\pi$. These oscillations lead to a modification of the DW equilibrium position due to the DW internal field ($H_{\rm int}(\phi)$), which exerts an additional torque on the DW in order to restore the equilibrium angle. As previously commented, if the amplitude of these oscillations is large enough, the DW gets depinned.  From Eq.~(\ref{eq:q0}) we see that the new equilibrium position (and therefore the amplitude of the oscillations) depends on the DMI field, the value of the DW angle $\phi$ and the damping parameter.

In particular, damping has a twofold influence on this dynamics: one the one hand, it appears directly in Eq.~(\ref{eq:q0}), dividing the internal field, meaning that for the same deviation of $\phi$ from equilibrium, we have a stronger internal field for smaller damping. 
On the other hand, the second influence of damping is on the DW internal angle:} once the DW angle has deviated from equilibrium, the restoring torque due to DMI is proportional to the damping parameter (see Eq.~(\ref{eq:phi0})). Hence, a lower damping leads to lower restoring torque and a larger deviation of $\phi$ from equilibrium. The maximum deviation of $\phi$ from equilibrium ($\delta\phi=\phi_{\rm max}-\phi_0$) is plotted in Fig.~\ref{fig:Fig8}(b) as function of damping for $\mu_0H_z=40$ mT.   As expected, a lower damping leads to a larger deviation $\delta\phi$. }

\r{In this latter section,  the DW was set at rest close to the barrier and, therefore, the initial DW velocity is zero.  Nevertheless, one might wonder what happens when the DW reaches the barrier with a finite velocity. We simulated this case by placing the DW at an initial distance $d_1=200$ nm from the barrier. The depinning is further reduced in this case (see Supplementary Material~\cite{Note1}for more details). However, in the static simulations, the depinning field remains constant, independently from the velocity at which the DW reaches the barrier, meaning that the reduction of $H_{\rm dep}$ is again related to the DW precession. When the DW starts from $d_1$ it reaches the barrier precessing, thus with a higher deviation from its equilibrium angle, leading to a higher effect of the internal field.  }

\r{\subsection{Different DMI and pinning barriers}}

\begin{figure}[h]
\centering
\includegraphics[width=0.35\textwidth]{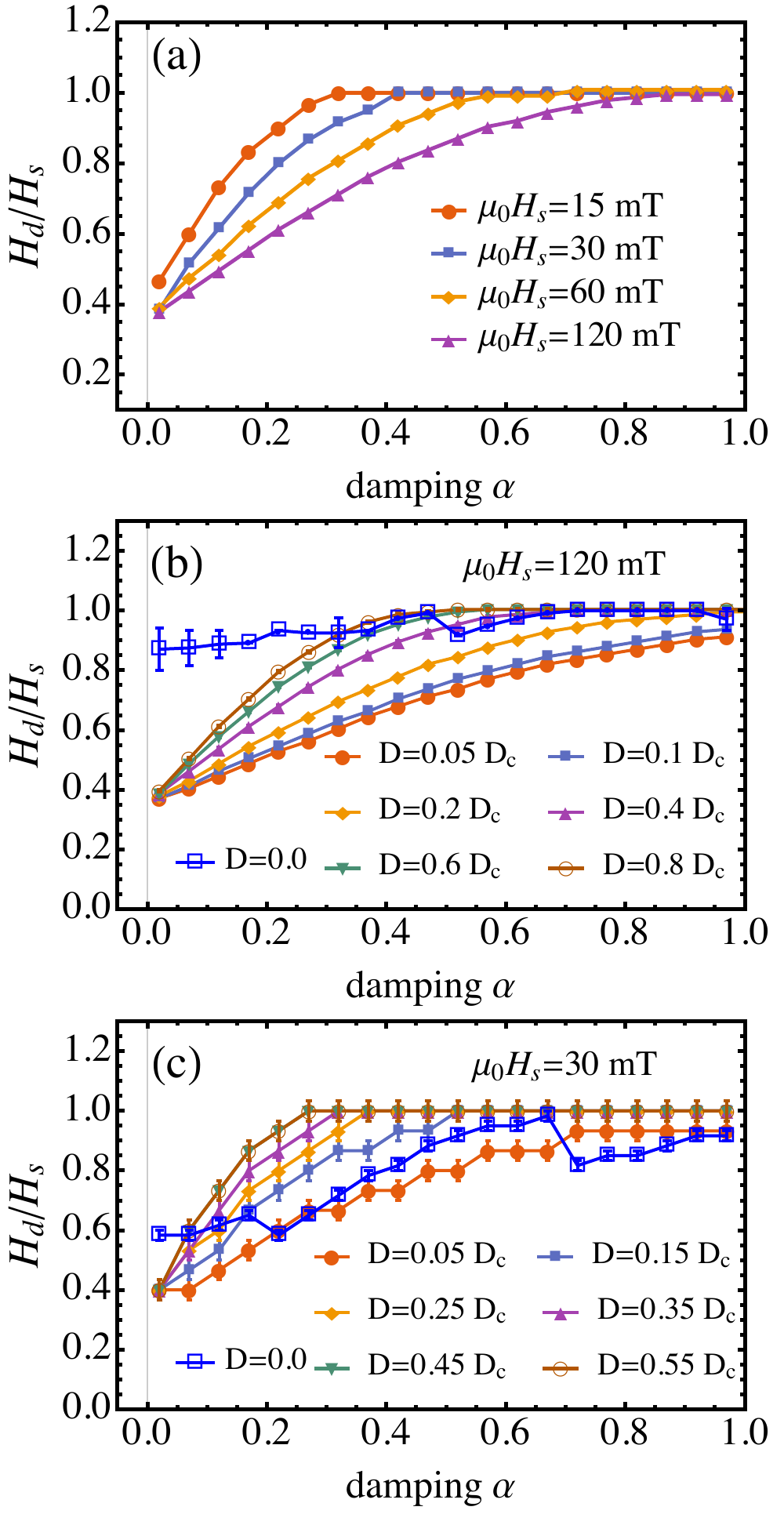}
\caption{(a) Dynamical $H_{\rm dep}$ as function of damping for different $H_s$ (disorder strength). (b) Dynamical $H_{\rm dep}$ as function of damping for different DMI constant and $\mu_0H_s=120$ mT. \r{(c)  Dynamical $H_{\rm dep}$ as function of damping for different DMI constant and $\mu_0H_s=30$ mT.}}
\label{fig:Fig10}
\end{figure}

\r{Finally, by using the 1D model it is possible to explore the dependence of $H_{\rm dep}$ on the pinning potential amplitude $H_s$ (related to the disorder strength) and on the DMI constant $D$.} The depinning field as function of damping for different values of $H_s$ is plotted in Fig.~\ref{fig:Fig10}(a). The reduction of $H_{\rm dep}$ is enhanced for larger values of $H_s$ (strong disorder). This is consistent with our explanation, since for strong disorder we need to apply larger fields \r{that lead to larger oscillations of $\phi$.}\\
Fig.~\ref{fig:Fig10}(b) represents the dynamical $H_{\rm dep}$ as function of damping for $\mu_0H_s=120$ mT and different DMI constants (expressed in term of the critical DMI constant $D_c=4\sqrt{AK_0}/\pi=3.9\ {\rm mJ/m^2}$)\footnote{For $D>D_c$, DW have negative energies and the systems spontaneously breaks into non-uniform spin textures.}. \r{In this case, the reduction of $H_{\rm dep}$ is enhanced for low DMI, until $D=0.05 D_c$, but a negligible reduction is observed for $D=0$. This non-monotonic behaviour can be explained by looking at the dependence of $\delta\phi$ and $H_{\rm int}$ on the DMI constant. 
}

\begin{figure}[h]
\includegraphics[width=0.35\textwidth]{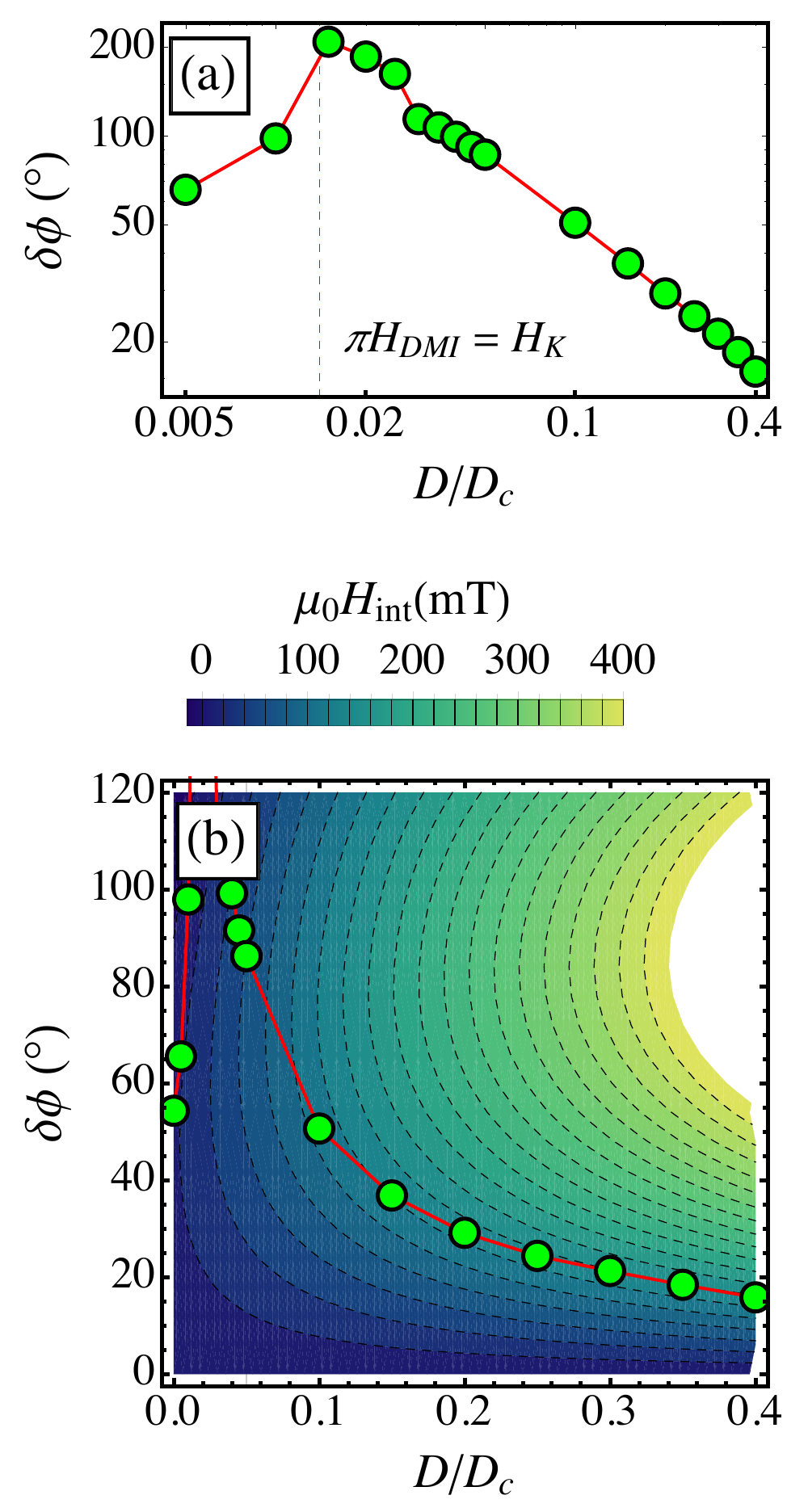}
\caption{(a) Max DW angle fluctuation $\delta\phi=\phi_{\rm max}-\phi_{\rm eq}$ as function of DMI for $\mu_0H_z=30$ mT. (b) Internal DW field $\mu_0 H_{\rm int}$ as function of DMI and $\delta\phi$. The green points correspond the max fluctuation plotted in (a). Note that the scale is logarithmic in (a). }
\label{fig:FigHint}
\end{figure}

\r{Fig.~\ref{fig:FigHint}(a) shows the maximum fluctuation $\delta\phi$ as function of DMI for $\mu_0H_z=30$ mT. $\delta\phi$ increases for low DMI  and it has a maximum at $\pi H_{\rm DMI}=H_K$, which in our case corresponds to $D=0.014D_c$. The increase of $\delta\phi$ for small values of $D$ is due to the smaller restoring torque in Eq.~(\ref{eq:phi0}). This holds until $\pi H_{\rm DMI}=H_K$, where shape anisotropy and DMI are comparable and they both affect the DW equilibrium configuration. As a consequence, the reduction of $H_{\rm dep}$ is enhanced by decreasing $D$ until $D\sim 0.014 D_c$, while it is reduced if $0<D<0.014 D_c$. Another contribution is given by the amplitude of the internal field, $H_{\rm int}$.  Fig.~\ref{fig:FigHint}(b) depicts $\mu_0H_{\rm int}$ as function of $\delta\phi$ and $D$. The maximum $\delta\phi$, obtained at $\mu_0H_z=30$ mT, is additionally marked in the plot. The internal field decreases with the DMI but this reduction is compensated by an increase in $\delta\phi$, which leads to an overall increase of $\mu_0 H_{\rm int}$, as discussed in the previous part.}
\r{However, at very low DMI, the internal field is dominated by shape anisotropy and, independently on the DW angle displacement, it is too small to have an effect on the depinning mechanism. Note, however, that the amplitude of $H_{\rm int}$ should be compared with the amplitude of the pinning barrier $H_s$.  Fig.~\ref{fig:Fig10}(b)  is calculated with $\mu_0H_s=120$ mT and the internal field, given by shape anisotropy ($H_K/2\sim 15$ mT), has indeed a negligible effect.  }
\r{However, larger effects are observed, in the case $D=0$, for smaller $H_s$, with reduction of $H_{\rm dep}$ up to $H_d/H_s\sim 0.6$, as shown in Fig.~\ref{fig:Fig10}(c), which is calculated with $\mu_0H_s=30$ mT. } \r{In other words, the reduction of the depinning field depends on the ratio between the pinning barrier and the internal DW field. } 

\begin{figure}[h]
\includegraphics[width=0.35\textwidth]{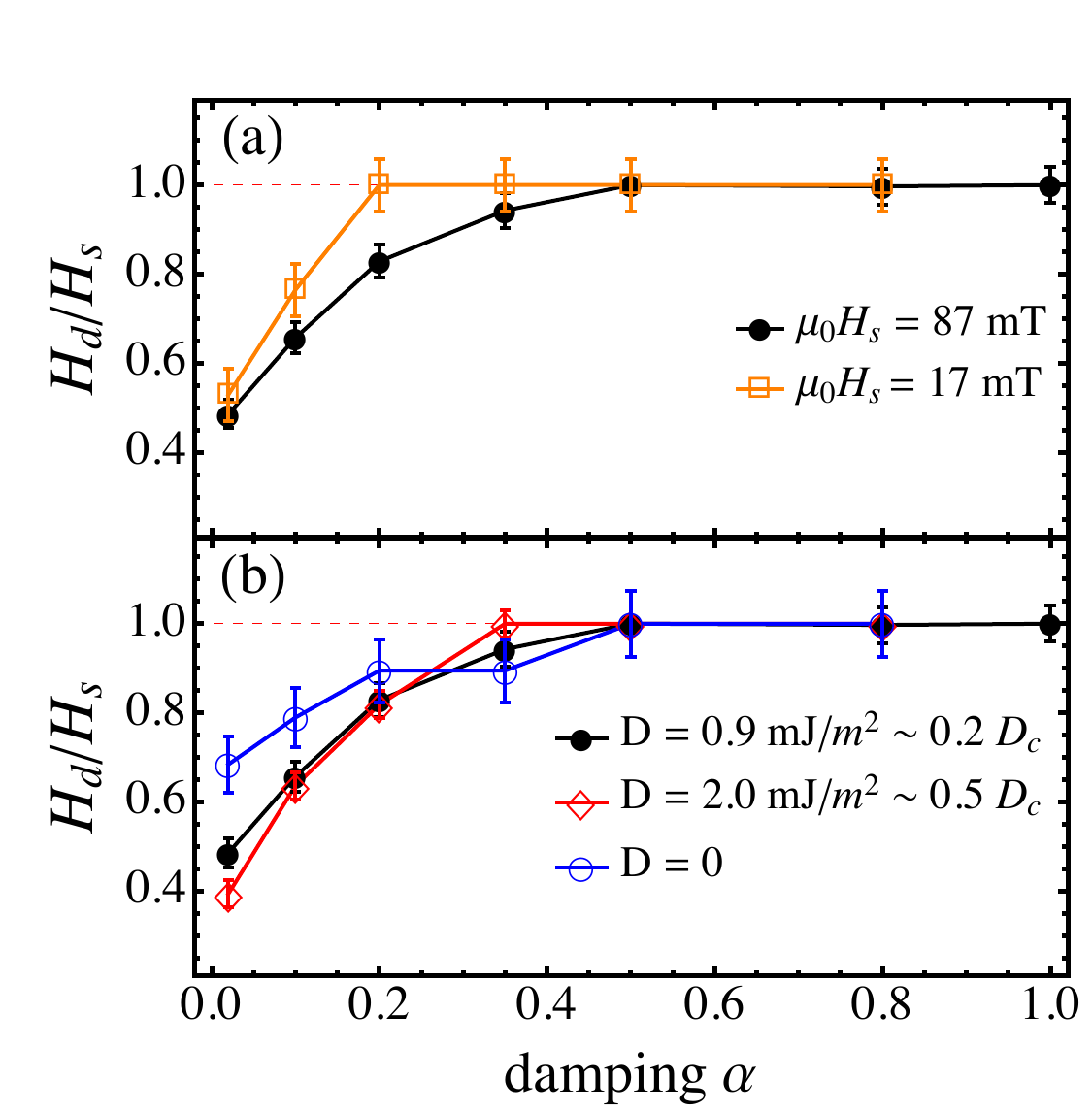}
\caption{(a) Dynamical $H_{\rm dep}$ as function of damping for different $H_s$ (disorder strength). (b) Dynamical $H_{\rm dep}$ as function of damping for different DMI constants.}
\label{fig:Fig11}
\end{figure}

Finally, it is interesting to see what happens for weaker disorder and different DMI in the system with grains. Fig.~\ref{fig:Fig11} shows the dynamical $H_{\rm dep}$, for different pinning potential and different DMI, obtained in the granular system. The results are in good agreement with what predicted by the 1D model for different disorder strengths.  However, we observe a smaller dependence on the DMI parameter. \r{This is due to two reasons: (1)~in the system with grains the static pinning barrier is $\mu_0H_s=87$ mT  and the dependence of the depinning field with DMI is smaller for smaller barriers, as shown in Fig.~\ref{fig:Fig10}(c). (2)~The DW motion in the granular system presents  the formation of VBL which might also contribute to the reduction of the depinning field. The mechanism is the same: a VBL is a non-equilibrium configuration for the DW (as a deviation of $\phi$ from equilibrium) that generates additional torques on the DW, which contribute to the DW depinning.}\\

\pagebreak
\clearpage
\section{Conclusions}
\label{sec:conclusions}

\b{To summarize, we have analysed the DW depinning field in a PMA sample with DMI and we found that $H_{\rm dep}$ decreases with the damping parameter with reductions up to $50\%$. This decrease is related to the DW internal dynamics and the finite size of the barrier: due to DW precession, the DW internal angle ($\phi$) deviates from equilibrium and triggers the \r{internal} DW field (\r{DMI and shape anisotropy}) which tries to restore its original value. At the same time, the internal field pushes the DW above its equilibrium position within the energy barrier. This mechanism leads to DW oscillations and, if the amplitude of the oscillations is higher than the barrier size, the DW gets depinned for a lower field. Deviations of $\phi$ from equilibrium and DW oscillations are both damping dependent and they are enhanced at low damping.}

In the system with grains the mechanism is the same but deviations from the internal DW equilibrium include the formation of VBL with more complex dynamics. The effect is enhanced for low DMI (providing that $\pi H_{\rm DMI}>H_{\rm K}$) and for stronger disorder since we need to apply larger external fields, which lead to larger DW oscillations.
These results are relevant both from a technological and theoretical point of view,  since they firstly suggest that a low damping parameter can lead to a lower $H_{\rm dep}$. Furthermore, they show that micromagnetic calculations of the depinning field, neglecting the DW precessional dynamics can provide only an upper limit for $H_{\rm dep}$, which could actually be lower due to the DW precessional dynamics. 
\\

\section{acknowledgement}
\label{sec:ack}
S.M. would like to thank K.~Shahbazi, C.H.~Marrows and J.~Leliaert for helpful discussions. 
This work was supported by Project WALL, FP7-
PEOPLE-2013-ITN 608031 from the European Commission, Project No. MAT2014-52477-C5-4-P from the Spanish government, and Project No. SA282U14 and SA090U16 from the Junta de Castilla y Leon.

\bibliographystyle{apsrev4-1}
\bibliography{library}

\appendix
\begin{widetext}
\section{Maximum torque and equilibrium state}
In this section we show in more detail how the maximum torque represents an indicator of the equilibrium state. 
Maximu  torque is defined as 
\begin{eqnarray}
\frac{\tau_{\rm max}}{\gamma_0}={\rm Max}\{-\frac{1}{1+\alpha^2}\m_i\times\HH_{{\rm eff},i}- \frac{\alpha}{1+\alpha^2}\m_i\times(\m_i\times\HH_{{\rm eff},i})\}=\frac{1}{\gamma_0}{\rm Max}\left(\frac{d\m_i}{dt}\right)\, ,
\end{eqnarray}
over all cells with label $i=\{1,...,N=N_x\cdot N_y\}$. MuMax3.9.3~\cite{Vansteenkiste2014} can provide this output automatically if selected.  We perform the same simulations as indicated in the main text, without any stopping condition, but simply running for $t=20$ ns.  Fig.~\ref{Fig:maxtorque}(a) shows the average $m_z$ component for $\alpha=0.2$ and $B_z=10$ mT, while Fig.~\ref{Fig:maxtorque}(b) depicts the corresponding maximum torque. We can see that, once the system has reached equilibrium, the maximum torque has dropped to a minimum value. The same results is obtained for different damping but the final maximum torque is different. \b{Numerically this value is never zero since it is limited by the code numerical precision and by the system parameters, in particular by damping.}

Fig.~\ref{Fig:maxtorque}(c) represents the maximum torque as function of applied field for different damping. The maximum torque is clearly independent on the applied field but depends on the damping value. Finally, Fig.~\ref{Fig:maxtorque}(d) shows the max torque as function of damping. The maximum torque decreases with damping and it saturates for $\alpha\geq 0.5$ since we have reached the minimum numerical precision of the code~\cite{Vansteenkiste2014}. For higher damping the maximum torque oscillates around this minimum sensibility value, as shown in the inset of Fig.~\ref{Fig:maxtorque}(d). The value obtained with these preliminary simulations is used to set a threshold $\epsilon (\alpha)$ for the depinning field simulations  in order to identify when the system has reached an equilibrium. Furthermore, additional tests were performed, without putting any max torque condition, but simply running the simulations for a longer time ($t=80,160$ ns) and calculating the depinning field in order to ensure that the results obtained with these two method were consistent, i.e., that we have actually reached an equilibrium state with the maximum torque condition. 
\begin{figure}[h]
\includegraphics[width=0.4\textwidth]{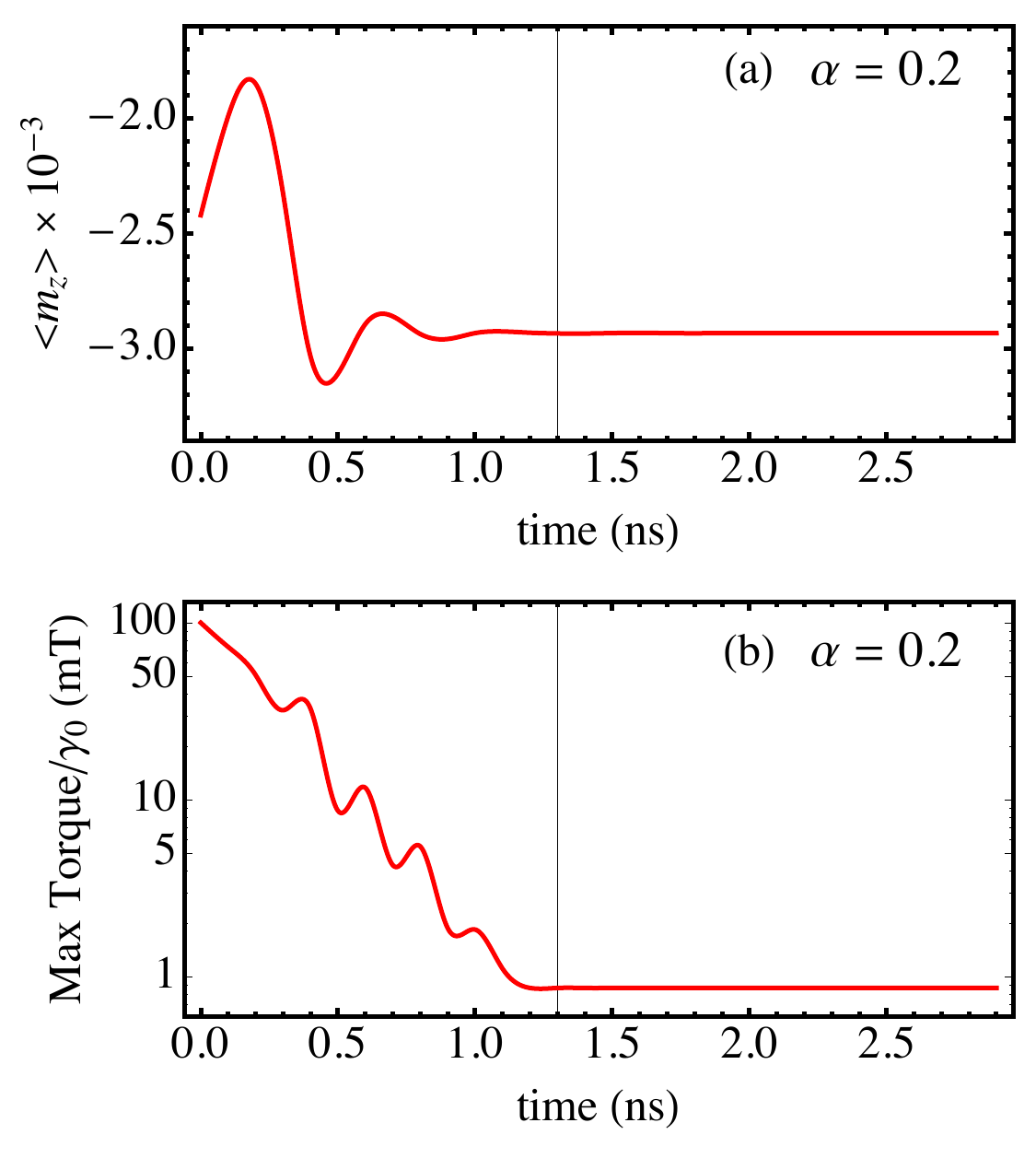}\qquad\qquad \includegraphics[width=0.4\textwidth]{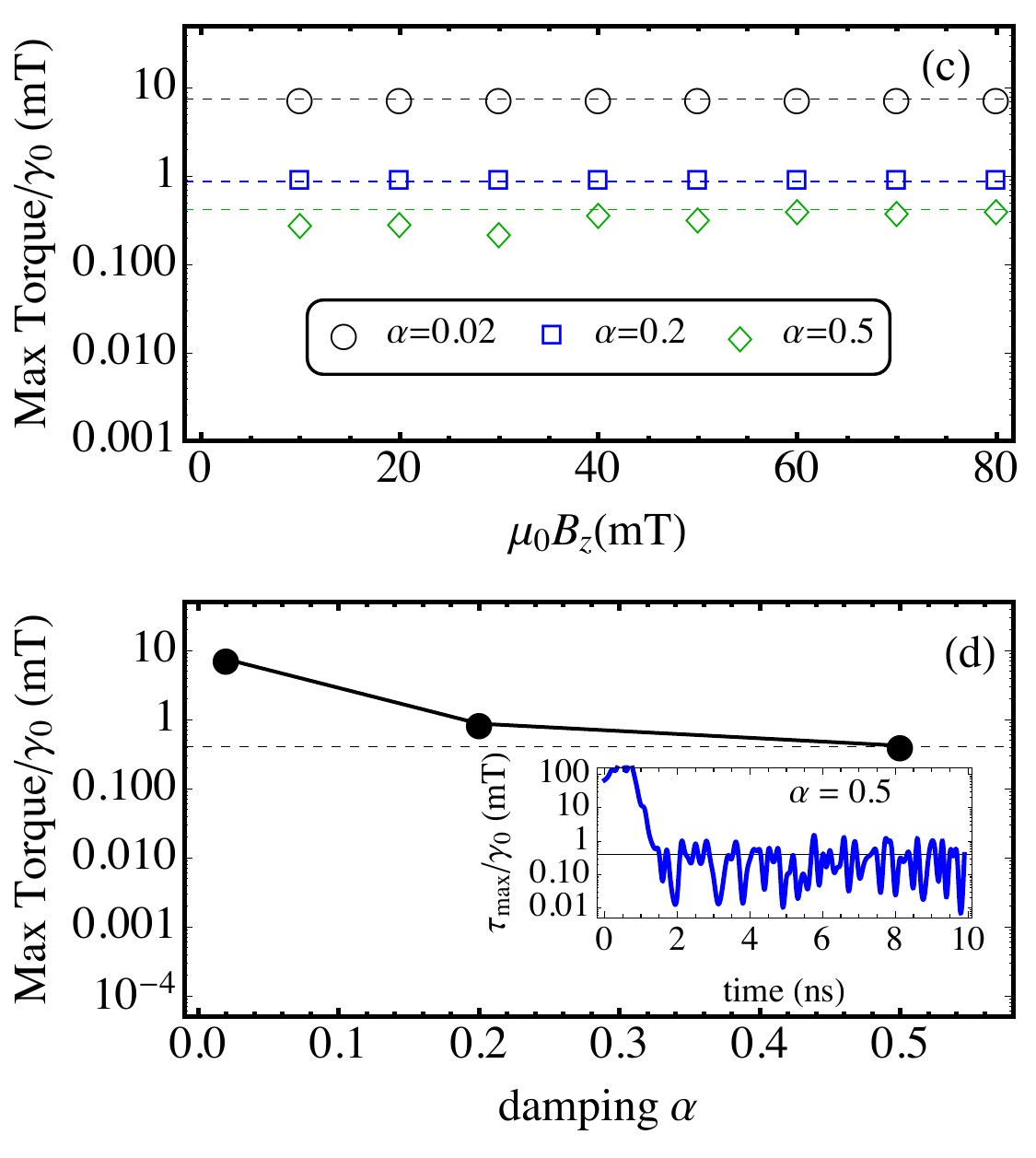}
\caption{(a) average $m_z$ as function of time. (b) Max torque/$\gamma_0$ ($\tau_{\rm max}$) as function of time.  $\tau_{\rm max}$ rapidly decreases when the system is at equilibrium. (c) Max torque as function of applied field for different damping. (d) Max torque at equilibrium as function of damping. The inset shows the max torque as function of time for $\alpha=0.5$.  }
\label{Fig:maxtorque}
\end{figure}
\pagebreak
\section{1D energy barrier}
As commented in the main text, the pinning field implemented in the 1D model simulations is obtained by using the shape of the DW energy profile derivative $\partial\sigma(x)/\partial x$ (being $x$ the DW position) and the amplitude of the depinning field obtained in the full micromagnetic simulations $H_s$ for the single barrier case. Namely
\begin{eqnarray}
H_{\rm dep}=H_s\left(\frac{\partial\sigma(x)}{\partial x}\right)_N\, ,
\label{eq:Hdep1}
\end{eqnarray}
where we recall that $N$ stands for the normalized value. This choice might sound unusual and needs to be justified. In fact, having the DW energy profile, the depinning field could be simply calculated as~\cite{Franken2011}
\begin{eqnarray}
H_{\rm dep}=\frac{1}{2\mu_0 M_s}\frac{\partial\sigma(x)}{\partial x}\, .
\label{eq:Hdep0}
\end{eqnarray}
This expression is derived by imposing that the derivative of the total DW energy $E(x)=2\mu_0M_sH_zx+\sigma(x)$ (Zeeman $+$ internal energy) must be always negative. However, in our case also $M_s(x)$ depends on the DW position and the results obtained with Eq.~\ref{eq:Hdep0} is different from the depinning field measured in the static single barrier simulations. For this reason we use Eq.~\ref{eq:Hdep1} which keep the correct barrier shape and has the measured static value.\\
Finally, we recall that equivalent results are obtained by using a simple Gaussian shape for the pinning field, meaning that the key point is the localized shape of the barrier, rather than its exact form.  
\section{Dynamical depinning for a moving Domain Wall}
In this section we show the results for the dynamical depinning field when the DW is placed at an initial distance of $d_1=200$ nm from the barrier. In this way the DW hits the pinning with an initial velocity. The $d_0$ case corresponds to the DW at rest relaxed just before the barrier and extensively analysed in the main text. Also for this configuration we performed static and dynamic simulations, neglecting or including the DW precessional dynamics respectively. The depinning field for the $d_1$ case is further reduces at small damping, reaching $H_d/H_s\sim 0.08 $ ($H_d=9$ mT and $H_s=120$ mT) at $\alpha=0.02$. Nevertheless, the depinning field remains constant in the static simulations independently on the velocity at which the DW hits the barrier. This suggests that, rather than related to the DW velocity, the reduction is again related to the DW precession. When the DW starts from $d_1$ it reaches the barrier precessing, thus with a higher displacement from its equilibrium angle, leading to a higher effect of the internal field.

\begin{figure}[h]
\includegraphics[width=0.4\textwidth]{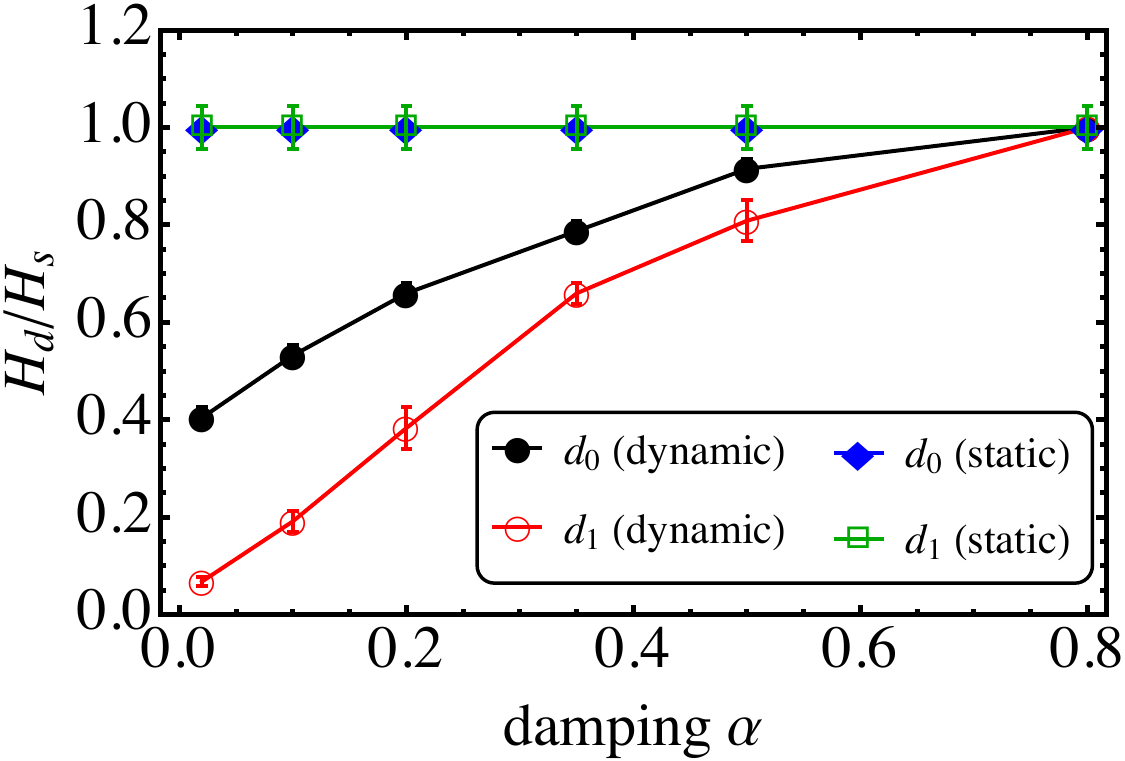}
\caption{Dynamical depinning field as function of damping for static and dynamic simulations for  the $d_0$ and $d_1$ cases. }
\label{Fig:Hd_farDW}
\end{figure}
\end{widetext}

\end{document}